\documentclass[12pt,letterpaper]{article}

\usepackage{hyperref} %%% note: automatically loads "color", use "dvips -z"
\usepackage{graphicx}
\usepackage{cite}
\begin{document}

%%%%%%%%%%%%%%%%%%%%%%%%%%%%%%%%%%%%%%%%%%%%%%%%%%%%%%%%%%%%%%%%%%%%%%%%%%%%%%%

\begin{center}
{\LARGE\bf The ratio $m_c/m_s$ with Wilson fermions}
\end{center}
\vspace{5pt}

\begin{center}
{\large\bf Stephan D\"urr$\,^{1,2}$}
\,\,\,and\,\,\,
{\large\bf Giannis Koutsou$\,^{3}$}
\\[10pt]
${}^1${\sl Bergische Universit\"at Wuppertal, Gau{\ss}stra{\ss}e\,20,
42119 Wuppertal, Germany}\\
${}^2${\sl J\"ulich Supercomputing Center, Forschungszentrum J\"ulich,
52425 J\"ulich, Germany}\\
${}^3${\sl Cyprus Institute, CaSToRC, 20 Kavafi Street, Nicosia 2121, Cyprus}
\end{center}
\vspace{5pt}

\begin{abstract}
\noindent
We determine the quark mass ratio $m_c/m_s$ on the lattice, using Wilson-type
fermions. Configurations with $N_f\!=\!2$ dynamical clover-improved fermions by
the QCDSF collaboration are used, which were made available through the ILDG.
In the valence sector we use a sophisticated, mass-independently
$O(a)$-improved Wilson-type action with small cut-off effects even in the charm
mass region. After an extrapolation to the physical pion mass, to zero lattice
spacing and to infinite box volume, we find $m_c/m_s=11.27(30)(26)$.
\end{abstract}
\vspace{5pt}

%\phantom{PACS: 12.38.Gc, 14.65.-q}

%%%%%%%%%%%%%%%%%%%%%%%%%%%%%%%%%%%%%%%%%%%%%%%%%%%%%%%%%%%%%%%%%%%%%%%%%%%%%%%

\newcommand{\pad}{\partial}
\newcommand{\hqu}{\hbar}
\newcommand{\ovr}{\over}
\newcommand{\til}{\tilde}
\newcommand{\pri}{^\prime}
\renewcommand{\dag}{^\dagger}
\newcommand{\<}{\langle}
\renewcommand{\>}{\rangle}
\newcommand{\gaf}{\gamma_5}
\newcommand{\nab}{\nabla}
\newcommand{\lap}{\triangle}
\newcommand{\dal}{{\sqcap\!\!\!\!\sqcup}}
\newcommand{\trc}{\mathrm{tr}}
\newcommand{\Trc}{\mathrm{Tr}}
\newcommand{\Mpi}{M_\pi}
\newcommand{\Fpi}{F_\pi}
\newcommand{\Mka}{M_K}
\newcommand{\Fka}{F_K}
\newcommand{\Met}{M_\et}
\newcommand{\Fet}{F_\et}
\newcommand{\Mss}{M_{\bar{s}s}}
\newcommand{\Fss}{F_{\bar{s}s}}
\newcommand{\Mcs}{M_{\bar{c}s}}
\newcommand{\Fcs}{F_{\bar{c}s}}
\newcommand{\Mcc}{M_{\bar{c}c}}
\newcommand{\Fcc}{F_{\bar{c}c}}

\newcommand{\al}{\alpha}
\newcommand{\be}{\beta}
\newcommand{\ga}{\gamma}
\newcommand{\de}{\delta}
\newcommand{\ep}{\epsilon}
\newcommand{\ve}{\varepsilon}
\newcommand{\ze}{\zeta}
\newcommand{\et}{\eta}
\renewcommand{\th}{\theta}
\newcommand{\vt}{\vartheta}
\newcommand{\io}{\iota}
\newcommand{\ka}{\kappa}
\newcommand{\la}{\lambda}
\newcommand{\rh}{\rho}
\newcommand{\vr}{\varrho}
\newcommand{\si}{\sigma}
\newcommand{\ta}{\tau}
\newcommand{\ph}{\phi}
\newcommand{\vp}{\varphi}
\newcommand{\ch}{\chi}
\newcommand{\ps}{\psi}
\newcommand{\om}{\omega}

\newcommand{\psb}{\bar{\psi}}
\newcommand{\etb}{\bar{\eta}}
\newcommand{\psh}{\hat{\psi}}
\newcommand{\eth}{\hat{\eta}}
\newcommand{\psd}{\psi^{\dagger}}
\newcommand{\etd}{\eta^{\dagger}}
\newcommand{\qh}{\hat{q}}
\newcommand{\kh}{\hat{k}}

\newcommand{\bdm}{\begin{displaymath}}
\newcommand{\edm}{\end{displaymath}}
\newcommand{\bea}{\begin{eqnarray}}
\newcommand{\eea}{\end{eqnarray}}
\newcommand{\beq}{\begin{equation}}
\newcommand{\eeq}{\end{equation}}

\newcommand{\mr}{\mathrm}
\newcommand{\mb}{\mathbf}
\newcommand{\Nf}{N_{\!f}}%{{N_{\!f}}}
\newcommand{\Nc}{N_{ c }}%{{N_{ c }}}
\newcommand{\Nt}{N_{ t }}%{{N_{ t }}}
\newcommand{\ri}{\mr{i}}
\newcommand{\DW}{D_\mr{W}}
\newcommand{\Dov}{D_\mr{ov}}
\newcommand{\Dst}{D_\mr{st}}
\newcommand{\Dke}{D_\mr{ke}}
\newcommand{\Dovm}{D_{\mr{ov},m}}
\newcommand{\Dstm}{D_{\mr{st},m}}
\newcommand{\Dkem}{D_{\mr{ke},m}}
\newcommand{\MeV}{\,\mr{MeV}}
\newcommand{\GeV}{\,\mr{GeV}}
\newcommand{\fm}{\,\mr{fm}}
\newcommand{\MSbar}{\overline{\mr{MS}}}

\newcommand{\tim}{\!\times\!}
\newcommand{\bull}{\bullet}

\def\lsim{\raise0.4ex\hbox{$<$\kern-0.78em\raise-0.9ex\hbox{$\sim$}}}
\def\gsim{\raise0.4ex\hbox{$>$\kern-0.78em\raise-0.9ex\hbox{$\sim$}}}

\hyphenation{topo-lo-gi-cal simu-la-tion theo-re-ti-cal mini-mum con-tinu-um}

%%%%%%%%%%%%%%%%%%%%%%%%%%%%%%%%%%%%%%%%%%%%%%%%%%%%%%%%%%%%%%%%%%%%%%%%%%%%%%%

\section{Introduction}

%%%%%%%%%%%%%%%%%%%%%%%%%%%%%%%%%%%%%%%%%%%%%%%%%%%%%%%%%%%%%%%%%%%%%%%%%%%%%%%

Quark masses are among the fundamental parameters of the Standard Model of
particle physics.
As they cannot be measured directly, their determination involves a substantial
amount of theory -- for decades uncertainties have been hard to estimate and
error-bars were large \cite{Nakamura:2010zzi}.
In recent years Lattice QCD has made enormous progress at pinning quark masses
down with a few-percent accuracy; see e.g.\ \cite{Colangelo:2010et} for a
summary.
For \emph{ratios} of quark masses the situation is even better, since in this
case no lattice-to-continuum matching factor (whose accurate determination
represents one of the most demanding steps in such a computation) is needed.

The charm-to-strange quark mass ratio $m_c/m_s$ (which is scheme and scale
independent) is of direct phenomenological relevance \cite{Rosner:2010ak}.
It has been determined by HPQCD \cite{Davies:2009ih} and ETM
\cite{Blossier:2010cr}.
Both collaborations use lattice formulations with small cut-off effects even in
the charm quark mass region, albeit with isospin (or taste) symmetry breaking,
i.e.\ the pions are non-degenerate, in spite of a single $m_q$ being used, an
effect which disappears $\propto\!a^2$ with $a$ the lattice spacing.
By contrast unimproved or $O(a)$-improved Wilson fermions avoid such effects,
at the price of having comparatively larger cut-off effects (see App.\,A of
\cite{Colangelo:2010et} for a discussion).

In \cite{Durr:2010ch} we constructed a Brillouin-improved Wilson action which
was claimed to show small cut-off effects without isospin breaking, thus
allowing for a one-to-one identification between lattice and continuum flavor.
The latter feature is important, as isospin breaking effects require a more
involved analysis, rendering it less transparent.
Here we test the smallness of the cut-off effects by calculating the ratio
$m_c/m_s$ in this formulation (with tree-level clover improvement and one step
of link smearing) in the valence sector (for $s$ and $c$).
The lattices with 2 degenerate dynamical flavors (for $u$ and $d$) are
provided by the QCDSF collaboration.
The remainder of this article describes how we calculate the ratio on each
ensemble, and how we remove the lattice artefacts to find the physical value
of $m_c/m_s$.
We end with an illustration of how this ratio may be used, together with a
precise $m_c$ input, to yield a robust estimate of $m_s$.

%%%%%%%%%%%%%%%%%%%%%%%%%%%%%%%%%%%%%%%%%%%%%%%%%%%%%%%%%%%%%%%%%%%%%%%%%%%%%%%

\section{\texorpdfstring{Strategy to compute $m_c/m_s$ on each ensemble}{txt1}}

%%%%%%%%%%%%%%%%%%%%%%%%%%%%%%%%%%%%%%%%%%%%%%%%%%%%%%%%%%%%%%%%%%%%%%%%%%%%%%%

Our goal is to compute the quark mass ratio $m_c/m_s$ with controlled
systematics.
We follow a two-step procedure.
In the first step we tune, for each ensemble, the bare mass parameter $\ka$ of
our action (see \cite{Durr:2010ch}) to the physical strange or charm quark mass
and evaluate $m_c/m_s$ on that ensemble.
In the second step we eliminate the lattice artefacts by means of a global fit.

Our strategy to compute $m_c/m_s$ on a given ensemble can be summarized as
follows:
\begin{enumerate}
\itemsep-2pt
%\item
%Tune $\ka_s$ to the point where $M_{\et_s}^2/(M_\ph^2\!-\!M_{\et_s}^2)$
%takes the value $0.8265$.
%\item
%Tune $\ka_c$ to the point where $M_{\et_c}^2/(M_{J/\ps}^2\!-\!M_{\et_c}^2)$
%takes the value $12.535$.
%\item
%Tune $\ka_c$ and $\ka_s$ at the same time such that $M_{\et_c}/M_{D_s}$ and
%$M_{D_s}/M_{\et_s}$ take the values $1.514$ and $2.870$, respectively.
%\item
%Tune $\ka_s$ and $\ka_c$ at the same time such that
%$M_{\et_s}^2/(M_{D_s^*}^2\!-\!M_{D_s}^2)$ and
%$M_{\et_c}^2/(M_{D_s^*}^2\!-\!M_{D_s}^2)$ take their physical values of
%$0.80138$ and $15.135$, respectively \cite{Nakamura:2010zzi}.
\item
Tune $\ka_s$ and $\ka_c$ at the same time such that
$M_{\et_s}^2/(M_{D_s^*}^2\!-\!M_{D_s}^2)$ and
$(2M_{D_s}^2\!-\!M_{\et_s}^2)/(M_{D_s^*}^2\!-\!M_{D_s}^2)$ take their physical
values of $0.80138$ and $12.402$, respectively \cite{Nakamura:2010zzi}.
These numbers build on $M_{\et_s}\!=\!0.6858(8)\GeV$ for the quark-line
connected state, which follows via $(2\Mka^2\!-\!\Mpi^2)^{1/2}$ with
SU(2)-symmetric values of $\Mka,\Mpi$ from \cite{Colangelo:2010et}, or from a
direct computation \cite{Davies:2009ih}.
\item
Determine for either tuned $\ka$ the PCAC quark mass, and form the ratio
$r\!=\!m_c^\mr{PCAC}/m_s^\mr{PCAC}$.
In this step $m_s^\mr{PCAC}$ is determined from the connected $\bar{s}s$
correlator, while $m_c^\mr{PCAC}$ follows from the $\bar{c}s$ correlator
together with the strange mass determined before (see below).
\end{enumerate}
%%%
%With Wilson-type fermions, it is important to notice that sea
%flavors affect the renormalization of valence flavor properties.
%For the PCAC quark mass the renormalization pattern is
%\cite{Bhattacharya:2005rb}
%\beq
%m_j^\mr{AWI}=\frac{Z_A}{Z_P}m_j^\mr{PCAC}
%\Big[
%1+(b_A\!-\!b_P)am_j^\mr{W}+(\bar{b}_A\!-\!\bar{b}_P)a\,\mr{Tr}(M)+O(a^2)
%\Big]
%\label{awi_bhattacharya}
%\eeq
%where $j$ denotes a flavor and $m^\mr{W}$ is the Wilson quark mass.
%In a ratio of AWI quark masses
%\beq
%\frac{m_j^\mr{AWI}}{m_k^\mr{AWI}}=\frac{m_j^\mr{PCAC}}{m_k^\mr{PCAC}}
%\Big[
%1+(b_A\!-\!b_P)a(m_j^\mr{W}\!-\!m_k^\mr{W})+O(a^2)
%\Big]
%\label{awi_ratio}
%\eeq
%the dependence on the coefficients $\bar{b}_A$, $\bar{b}_P$ disappears.
%With our tree-level $O(a)$-improvement strategy things are even simpler, since
%$b_A\!-\!b_P\!=\!0$ implies that the ratio $r\!=\!m_c^\mr{PCAC}/m_s^\mr{PCAC}$
%does not undergo renormalization at all.
%Formally, $b_A\!-\!b_P\!=\!O(g_0^2)$, and this means that our leading cut-off
%effects are $O(\al_\mr{s}a)$, where $\al_\mr{s}\!=\!g^2/(4\pi)$ is the strong
%coupling constant.
%Numerically, $a(m_c\!-\!m_s)\!\sim\!0.35$ at $a^{-1}\simeq2.8\GeV$ [see
%(\ref{qcdsf_spacings}) below].
%For unsmeared clover fermions typically $|b_A\!-\!b_P|=O(10^{-1})$, and for
%our action \cite{Durr:2010ch} it is reasonable to expect
%$|b_A\!-\!b_P|=O(10^{-2})$.
%Hence, upon ignoring the $b_A\!-\!b_P$ term in (\ref{awi_ratio}) and
%extrapolating the data with an $O(a^2)$ ansatz alone (which we won't do), one
%would ignore an effect of order $\lsim$1\%.
%%%
As a theoretical caveat let us remark that in general with Wilson-type fermions
the sea quarks affect the renormalization properties of the valence flavors.
For a bare PCAC quark mass \cite{Bhattacharya:2005rb}
\beq
m_j^\mr{AWI}=\frac{Z_A}{Z_P}m_j^\mr{PCAC}
\Big[
1+(b_A\!-\!b_P)am_j^\mr{W}+(\bar{b}_A\!-\!\bar{b}_P)a\,\mr{Tr}(M)+O(a^2)
\Big]
\label{awi_bhattacharya}
\eeq
where $m_j^\mr{W}$ is the Wilson mass of flavor $j$, and $M$ the quark mass
matrix.
The $Z_J$ with $J\!\in\!\{A,P\}$ are lattice-to-continuum matching factors,
while $b_J\!=\!1\!+\!O(\al_\mr{s})$, $\bar{b}_J\!=\!O(\al_\mr{s}^2)$ denote
improvement coefficients.
As we follow a tree-level improvement strategy (with $c_\mr{SW}\!=\!1$, see
\cite{Durr:2010ch}) the ratio
\beq
\frac{m_j^\mr{AWI}}{m_k^\mr{AWI}}=\frac{m_j^\mr{PCAC}}{m_k^\mr{PCAC}}
\Big[
1+O(\al_\mr{s}a)+O(a^2)
\Big]
\label{awi_ratio}
\eeq
is found to carry two types of cut-off effects.
As we shall see, the lack of knowledge which type would numerically dominate
creates a major source of systematic error on the final result.
%%%

Once $r\!=\!m_c^\mr{PCAC}/m_s^\mr{PCAC}$ is in hand for each ensemble, the
final answer follows through three more steps (which, in practice, will be
combined into a single global fit):
\begin{enumerate}
\itemsep-2pt
\setcounter{enumi}{2}
\item
Correct, for each ensemble, the value of $r$ for the effect of the finite
spatial volume $L^3$.
\item
Extrapolate, for each $\be$, the result of step 3 to
$\Mpi^\mr{phys}\!=\!134.8\MeV$ \cite{Colangelo:2010et} in the sea.
\item
Extrapolate the result of step 4 to the continuum, using an $O(\al_\mr{s}a)$ or
$O(a^2)$ ansatz.
\end{enumerate}
To test how reliably the systematic uncertainties are assessed, we will repeat
steps 3-5 for the control quantity $M_\ph^2/(M_{D_s^*}^2\!-\!M_{D_s}^2)$, whose
physical value is known.

%%%%%%%%%%%%%%%%%%%%%%%%%%%%%%%%%%%%%%%%%%%%%%%%%%%%%%%%%%%%%%%%%%%%%%%%%%%%%%%

\section{\texorpdfstring{Analysis details and final result for $m_c/m_s$}{txt2}}

%%%%%%%%%%%%%%%%%%%%%%%%%%%%%%%%%%%%%%%%%%%%%%%%%%%%%%%%%%%%%%%%%%%%%%%%%%%%%%%

We now give details of how we determine the ratio $m_c/m_s$ on each ensemble,
and how we eliminate the lattice artefacts by means of a global fit.

\begin{table}[!tb]
\centering
\small
\begin{tabular}{|c|cc|cccc|ccc|c|}
\hline
$\be$ & $\ka_\mr{sea}$ & box size &
$a\Mpi\!$\cite{Gockeler:2006jt,Bietenholz:2010az} &
$\!\Mpi[\!\MeV]\!$ & $\!L[\!\fm]\!$ & $\!\Mpi\!L\!$ &
$\!\Mpi[\!\MeV]\!$ & $\!a[\!\fm]\!$ & $\!L[\!\fm]\!$ &
use \\
\hline
%5.20 & 0.13420 & $16^3\tim32$ & 0.5847(12) &      & 1.33 &     &         &       &     &         \\
%\hline
 5.25 & 0.13460 & $16^3\tim32$ & 0.4932(10) & 1281 & 1.22 & 7.9 &  987(2) & 0.099 & 1.6 & $\bull$ \\ %$\circ$
% NOT & 0.13520 & $16^3\tim32$ & 0.3821(13) &  992 & 1.22 & 6.1 &  829(3) & 0.091 & 1.5 &         \\
      & 0.13575 & $24^3\tim48$ & 0.2556(06) &  664 & 1.82 & 6.1 &  597(1) & 0.084 & 2.0 & $\bull$ \\
      & 0.13600 & $24^3\tim48$ & 0.1849(---)&  480 & 1.82 & 4.4 &  ---    & ---   & --- & $\bull$ \\
\hline
 5.29 & 0.13500 & $16^3\tim32$ & 0.4206(09) & 1153 & 1.15 & 6.7 &  929(2) & 0.097 & 1.4 &         \\
%     & 0.13550 & $12^3\tim32$ & ???        &  ??? & 0.86 & ??? &  ???    & ???   & ??? &         \\
      & 0.13550 & $16^3\tim32$ & 0.3325(13) &  911 & 1.15 & 5.3 &  ---    & ---   & --- &         \\
      & 0.13550 & $24^3\tim48$ & 0.3270(06) &  896 & 1.73 & 7.8 &  769(2) & 0.089 & 2.0 & $\bull$ \\
%     & 0.13590 & $12^3\tim32$ & ???        &  ??? & 0.86 & ??? &  ???    & ???   & ??? &         \\
      & 0.13590 & $16^3\tim32$ & 0.2518(15) &  690 & 1.15 & 4.0 &  ---    & ---   & --- &         \\
      & 0.13590 & $24^3\tim48$ & 0.2395(05) &  656 & 1.73 & 5.7 &  591(2) & 0.084 & 1.9 & $\bull$ \\
      & 0.13620 & $24^3\tim48$ & 0.1552(06) &  425 & 1.73 & 3.7 &  395(3) & 0.080 & 1.9 & $\bull$ \\
      & 0.13632 & $24^3\tim48$ & 0.1106(12) &  303 & 1.73 & 2.7 &  ---    & ---   & --- & $\circ$ \\
      & 0.13632 & $32^3\tim64$ & 0.1075(09) &  295 & 2.30 & 3.4 &  337(3) & 0.077 & 2.5 & $\bull$ \\
      & 0.13632 & $40^3\tim64$ & 0.1034(08) &  283 & 2.88 & 4.1 &  ---    & ---   & --- &         \\
      & 0.13640 & $40^3\tim64$ & 0.0660(10) &  181 & 2.88 & 2.6 &  ---    & ---   & --- &         \\
%     & 0.13640 & $48^3\tim64$ & ???        &  ??? & 3.46 & ??? &  ???    & ???   & ??? &         \\
\hline
 5.40 & 0.13500 & $24^3\tim48$ & 0.4030(04) & 1325 & 1.44 & 9.7 & 1037(1) & 0.077 & 1.8 &         \\
      & 0.13560 & $24^3\tim48$ & 0.3123(07) & 1027 & 1.44 & 7.5 &  842(2) & 0.073 & 1.8 & $\bull$ \\
      & 0.13610 & $24^3\tim48$ & 0.2208(07) &  726 & 1.44 & 5.3 &  626(2) & 0.070 & 1.7 & $\bull$ \\
      & 0.13625 & $24^3\tim48$ & 0.1902(06) &  626 & 1.44 & 4.6 &  ---    & ---   & --- &         \\
      & 0.13640 & $24^3\tim48$ & 0.1538(10) &  506 & 1.44 & 3.7 &  432(3) & 0.068 & 1.6 & $\bull$ \\ %$\circ$
      & 0.13640 & $32^3\tim64$ & 0.1504(04) &  495 & 1.92 & 4.8 &  ---    & ---   & --- & $\bull$ \\
      & 0.13660 & $32^3\tim64$ & 0.0867(11) &  285 & 1.92 & 2.8 &  ---    & ---   & --- & $\bull$ \\
%     & 0.13660 & $48^3\tim64$ & ???        &  ??? & 2.89 & ??? &  ???    & ???   & ??? &         \\
\hline
\end{tabular}
\normalsize
\caption{\sl
Details of the QCDSF $\Nf\!=\!2$ lattices made available through the ILDG, with
$a\Mpi$ from \cite{Bietenholz:2010az} (in one case inferred from
\cite{Collins:2011mk}). The values of $\Mpi,L$ in the same block are based
on the scales (\ref{qcdsf_spacings}). For comparison we add another block with
information on $\Mpi$, $a$, $L$ from \cite{Gockeler:2006vi}.}
\label{tab:ensembles}
\end{table}

We use the $\Nf\!=\!2$ configurations by QCDSF
\cite{Gockeler:2006jt,Gockeler:2006vi,Bietenholz:2010az,Collins:2011mk} made
available through the ILDG \cite{Yoshie:2008aw}.
Since we measure dimensionless ratios, one might naively think that no scale
determination is needed.
However, in the extrapolation to the physical point a scale is required.
We will use \cite{Collins:2011mk}
\beq
%a[\mr{fm}]=0.083,0.076,0.072,0.060\quad\mr{at}\quad\be\!=\!5.20,5.25,5.29,5.40
a[\mr{fm}]=0.076,0.072,0.060\quad\mr{at}\quad\be\!=\!5.25,5.29,5.40
\label{qcdsf_spacings}
\eeq
for this purpose, but apart from the extrapolation this scale is not used.
Given the resources available to us, we select the 13 ensembles marked with a
bullet or circle in Tab.\,\ref{tab:ensembles} for analysis.
They cover a wide range of pion masses and box volumes (both in fm and in
$\Mpi L$ units), so that a controlled extrapolation to the physical pion mass
and infinite volume should be possible.

\begin{table}[!tb]
\centering
\small
\begin{tabular}{|c|ccc|cccc|}
\hline
$\be$ & $\ka_\mr{sea}$ & box size & confs & $1/\ka_s$ & $1/\ka_c$ & $O_3$ & $O_4$ \\
\hline
%5.20& 0.13420 & $16^3\tim32$ &         500 & 7.7943(25) & 8.965(25) & 13.35(23) & 2.303(45) \\
%\hline
%5.25 & 0.13460 & $16^3\tim32$ &         500 & 7.8313(25) & 8.787(26) & 13.34(29) & 2.270(63) \\
%     & 0.13575 & $24^3\tim48$ &         500 & 7.8496(16) & 8.601(20) & 12.90(13) & 2.054(54) \\
%     & 0.13600 & $24^3\tim48$ &         500 & 7.8530(16) & 8.558(17) & 12.76(16) & 2.141(53) \\
%\hline
%5.29 & 0.13550 & $24^3\tim48$ & 2$\cdot$400 & 7.8601(13) & 8.641(13) & 12.69(15) & 2.123(39) \\
%     & 0.13590 & $24^3\tim48$ & 2$\cdot$500 & 7.8632(15) & 8.574(16) & 12.68(15) & 2.089(42) \\
%     & 0.13620 & $24^3\tim48$ &         500 & 7.8645(14) & 8.518(13) & 12.83(20) & 2.036(51) \\
%     & 0.13632 & $24^3\tim48$ &         386 & 7.8615(17) & 8.477(17) & 13.41(25) & 2.031(57) \\
%     & 0.13632 & $32^3\tim64$ &         500 & 7.8653(11) & 8.485(12) & 12.61(10) & 1.990(36) \\
%\hline
%5.40 & 0.13560 & $24^3\tim48$ &         500 & 7.8818(14) & 8.499(14) & 13.07(21) & 2.167(64) \\
%     & 0.13610 & $24^3\tim48$ &         500 & 7.8850(14) & 8.458(12) & 12.71(24) & 2.064(54) \\
%     & 0.13640 & $24^3\tim48$ &         500 & 7.8844(15) & 8.410(15) & 12.73(24) & 2.114(71) \\
%     & 0.13640 & $32^3\tim64$ &         500 & 7.8864(11) & 8.413(12) & 12.16(15) & 1.948(42) \\
%     & 0.13660 & $32^3\tim64$ &         500 & 7.8862(10) & 8.396(10) & 12.40(12) & 1.958(37) \\
%\hline
5.25 & 0.13460 & $16^3\tim32$ & 2$\cdot$500 & 7.8310(18) & 8.793(19) & 13.48(22) & 2.220(54) \\
     & 0.13575 & $24^3\tim48$ & 2$\cdot$500 & 7.8504(12) & 8.612(15) & 12.87(11) & 2.081(38) \\
     & 0.13600 & $24^3\tim48$ & 2$\cdot$500 & 7.8520(11) & 8.548(11) & 12.89(12) & 2.114(36) \\
\hline
5.29 & 0.13550 & $24^3\tim48$ & 2$\cdot$400 & 7.8601(13) & 8.641(13) & 12.69(16) & 2.123(39) \\
     & 0.13590 & $24^3\tim48$ & 2$\cdot$500 & 7.8632(15) & 8.574(16) & 12.68(16) & 2.089(42) \\
     & 0.13620 & $24^3\tim48$ & 2$\cdot$500 & 7.8635(09) & 8.502(09) & 12.85(14) & 2.056(34) \\
     & 0.13632 & $24^3\tim48$ &         386 & 7.8615(17) & 8.477(17) & 13.41(25) & 2.031(57) \\
     & 0.13632 & $32^3\tim64$ & 2$\cdot$500 & 7.8648(08) & 8.478(09) & 12.57(08) & 1.977(27) \\
\hline
5.40 & 0.13560 & $24^3\tim48$ & 2$\cdot$500 & 7.8823(10) & 8.503(10) & 13.01(15) & 2.163(46) \\
     & 0.13610 & $24^3\tim48$ & 2$\cdot$500 & 7.8859(10) & 8.463(09) & 12.55(17) & 2.073(37) \\
     & 0.13640 & $24^3\tim48$ & 2$\cdot$500 & 7.8842(11) & 8.403(12) & 12.66(19) & 2.041(51) \\
     & 0.13640 & $32^3\tim64$ & 2$\cdot$500 & 7.8864(08) & 8.417(08) & 12.25(11) & 1.962(28) \\
     & 0.13660 & $32^3\tim64$ & 2$\cdot$500 & 7.8862(08) & 8.397(08) & 12.42(11) & 1.955(29) \\
\hline
\end{tabular}
\normalsize
\caption{\label{tab:results}\sl Tuned kappas of the Brillouin operator and
final $O_{3,4}$ for each ensemble. Usually $500$ configs were downloaded;
in most cases inversions were performed on more than one timeslice.}
\vspace*{-2mm}
\end{table}

On a given ensemble, for a few mass parameters $1/\ka_s,1/\ka_c$,
%%%
%four observables are determined,
%$O_1\!=\!M_{P,\bar{s}s}^2/(M_{V\bar{c}s}^2\!-\!M_{P\bar{c}s}^2)$,
%$O_2\!=\!M_{P,\bar{c}c}^2/(M_{V\bar{c}s}^2\!-\!M_{P\bar{c}s}^2)$, and two
%versions [to be dubbed $O_3,O_4$] of $m_c^\mr{PCAC}/m_s^\mr{PCAC}$.
%For each observable a spline interpolation in $1/\ka_s$ and $1/\ka_c$ is
%constructed.
%The target value $O_1\!\equiv\!0.80138$ defines a line in the
%$(1/\ka_s,1/\ka_c)$ plane, and the same holds true for $O_2\!\equiv\!15.135$.
%The point where these two lines intersect defines the tuned set
%$(1/\ka_s^*,1/\ka_c^*)$, and the value of $m_c/m_s$ at that point is the
%desired ratio on that ensemble.
%The spacing in $1/\ka_s$ and $1/\ka_c$ is chosen sufficiently narrow for
%the uncertainty in the interpolation being negligible.
%A complication with Wilson fermions is that the definition of $r$ is not unique.
%The PCAC recipe $am_i\!+\!am_j=\<\pad A_4(t)P(0)\>/\<P(t)P(0)\>$ yields the sum
%of the masses of two flavors $i,j$ involved in either correlator.
%Hence, after determining $2m_s^\mr{PCAC}$ from the $\bar{s}s$ correlators,
%one may determine $m_c^\mr{PCAC}\!+\!m_s^\mr{PCAC}$ from the $\bar{c}s$
%correlators [and form $r'\!=\!O_3(1/\ka_s^*,1/\ka_c^*)$], or one may determine
%$2m_c^\mr{PCAC}$ from the $\bar{c}c$ correlators [and form
%$r''\!=\!O_4(1/\ka_s^*,1/\ka_c^*)$].
%In Tab.\,\ref{tab:results} we list the tuned values $(1/\ka_s^*,1/\ka_c^*)$ for
%each ensemble, along with $r'$ and $r''$.
%This ambiguity is an $O(\al_\mr{s}a)$ or $O(a^2)$ cut-off effect, which is
%supposed to disappear with the continuum extrapolation.
%%%
we determine the correlators of four mesons (the pseudoscalar and vector with
$\bar{c}s$ and $\bar{s}s$ flavor content).
From these we form the observables
$O_1\!=\!M_{P\bar{s}s}^2/(M_{V\bar{c}s}^2\!-\!M_{P\bar{c}s}^2)$,
$O_2\!=\!(2M_{P\bar{c}s}^2-M_{P\bar{s}s}^2)/(M_{V\bar{c}s}^2\!-\!M_{P\bar{c}s}^2)$,
and $O_3\!=\!(2m_{cs}\!-\!m_{ss})/m_{ss}$, where $m_{ij}$ denotes the average
of the PCAC masses with flavor $i$ and $j$, based on the improved symmetric
derivative
$\bar\pad\ph(t)=[\ph(t\!-\!2)-8\ph(t\!-\!1)+8\ph(t\!+\!1)-\ph(t\!+\!2)]/12$.
For each observable a spline interpolation in $1/\ka_s$ and $1/\ka_c$ is
constructed.
The target value $O_1\!\equiv\!0.80138$ defines a line in the
$(1/\ka_s,1/\ka_c)$ plane, and the same holds true for $O_2\!\equiv\!12.402$.
The point where these two lines intersect defines the tuned set
$(1/\ka_s^*,1/\ka_c^*)$, and the value of $O_3$ at this point is the
desired ratio $r$ on that ensemble.
The spacing in $1/\ka_s$ and $1/\ka_c$ is chosen sufficiently narrow so that
the uncertainty due to the interpolation is completely negligible.
Since all of this is done inside a jackknife, the jitter of the crossing point
is fully propagated into the statistical error of the tuned $r$, as listed in
Tab.\,\ref{tab:results}.
For an illustration see \cite{arXiv:1111.2577}.
%%%

Finally, we wish to correct for the systematic effect that the finite lattice
spacing ($a\!>\!0$), the larger-than-physical pion mass
($\Mpi\!>\!\Mpi^\mr{phys}$) and the finite spatial volume ($L^3\!<\!\infty$)
have on the measured $m_c/m_s$, by means of a global fit to our dataset.
For each artefact, we invoke an extrapolation formula which is consistent with
both theoretical expectations and the data.
We shall consider several (reasonable) options for each effect, and treat the
spread of these as the systematic error of the final result.
The dominant cut-off effects may be proportional to $\al_\mr{s}a$ (what theory
suggests) or proportional to $a^2$ (what empirical evidence seems to prefer
\cite{Durr:2010ch}).
In the range of interest the dependence on $m_{u,d}^\mr{sea}$ may be a
quadratic or cubic function of $\Mpi$.
Finite volume effects may be proportional to
$K_1(\Mpi L)/(\Mpi L)\!\sim\!\exp(-\Mpi L)/(\Mpi L)^{3/2}$ times
$\Mpi^2/(4\pi\Fpi)^2$ (as in the case of $\Mpi(L)/\Mpi\!-\!1$), or just
proportional to $1/L^3$ (as frequently used in the old lattice literature).
By combining these forms we arrive at the 8 ans\"atze
\beq
r^{(i,j,k)}(a,\Mpi,L)=b\cdot
\Big[1\,+\,
c^{(i)}f^{(i)}(a)+
d^{(j)}g^{(j)}(\Mpi)+
e^{(k)}h^{(k)}(\Mpi,L)
\Big]
\label{ansatz}
\eeq
with $i,j,k\in\{1,2\}$, where $f^{(1)}\!=\!\al_\mr{s}a$, $f^{(2)}\!=\!a^2$,
$g^{(1)}\!=\!\Mpi^2$, $g^{(2)}\!=\!\Mpi^3$,
$h^{(1)}\!=\!\sqrt{\Mpi/L^3}\exp(-\Mpi L)$, $h^{(2)}\!=\!1/L^3$.
Note that this is the first time that we make use of the auxiliary scales
(\ref{qcdsf_spacings}); here we need to invoke them, since the coefficients
$c^{(i)},d^{(j)},e^{(k)}$ are dimensionful quantities.

The last point to be discussed is which ensembles are included in the fit.
It turns out that the ensemble 5.29\_0.13632\_$24^3\!\times\!48$ cannot be
described by any of the ans\"atze; once we drop it all versions of (\ref{ansatz})
yield consistently $\ch^2/\mr{d.o.f}\!\simeq\!1$.
For $r^{(i,j,k)}(0,\Mpi^\mr{phys},\infty)$ one finds $11.01(36)$, $11.02(32)$,
$11.24(33)$, $11.05(35)$, $11.39(26)$, $11.41(24)$, $11.60(25)$, $11.43(26)$,
respectively, where the errors are purely statistical.
To avoid underestimating the effect of the extrapolation, we need to include
the spread among these $8$ results as a source of systematic uncertainty.
This yields $m_c/m_s=11.27(30)(22)$, where the standard deviation of the
distribution is used as the systematic error.
Since some of the ensembles feature large pion masses and small volumes we
use the cuts $\Mpi\!<\!670,900\MeV$ and/or $L\!>\!1.4,1.7\fm$ to check for
any additional systematic uncertainties.
This yields six independent options for our dataset.
The center of this enlarged distribution (from the $8\!\cdot\!6\!=\!48$
analyses) is lower than the central value mentioned above, amounting to an
additional systematic uncertainty of $0.14$ which we add in quadrature to the
previous one.
This yields our final result
\beq
m_c/m_s=11.27(30)(26)
\label{result}
\eeq
in the continuum, at the physical mass point, and in infinite volume.

\begin{figure}[!tbp]
\centering
%%%
\includegraphics[width=9cm]{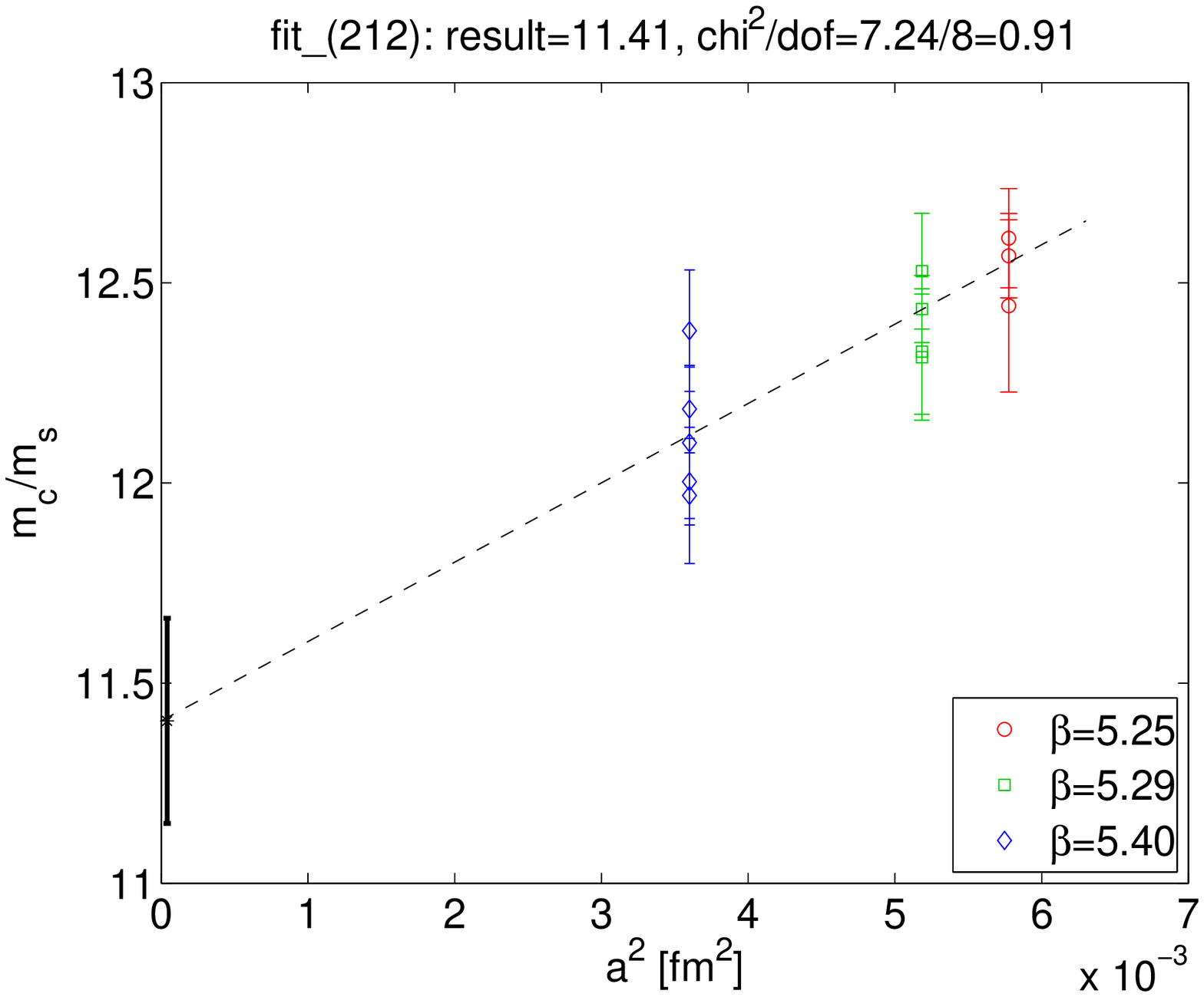}
\includegraphics[width=9cm]{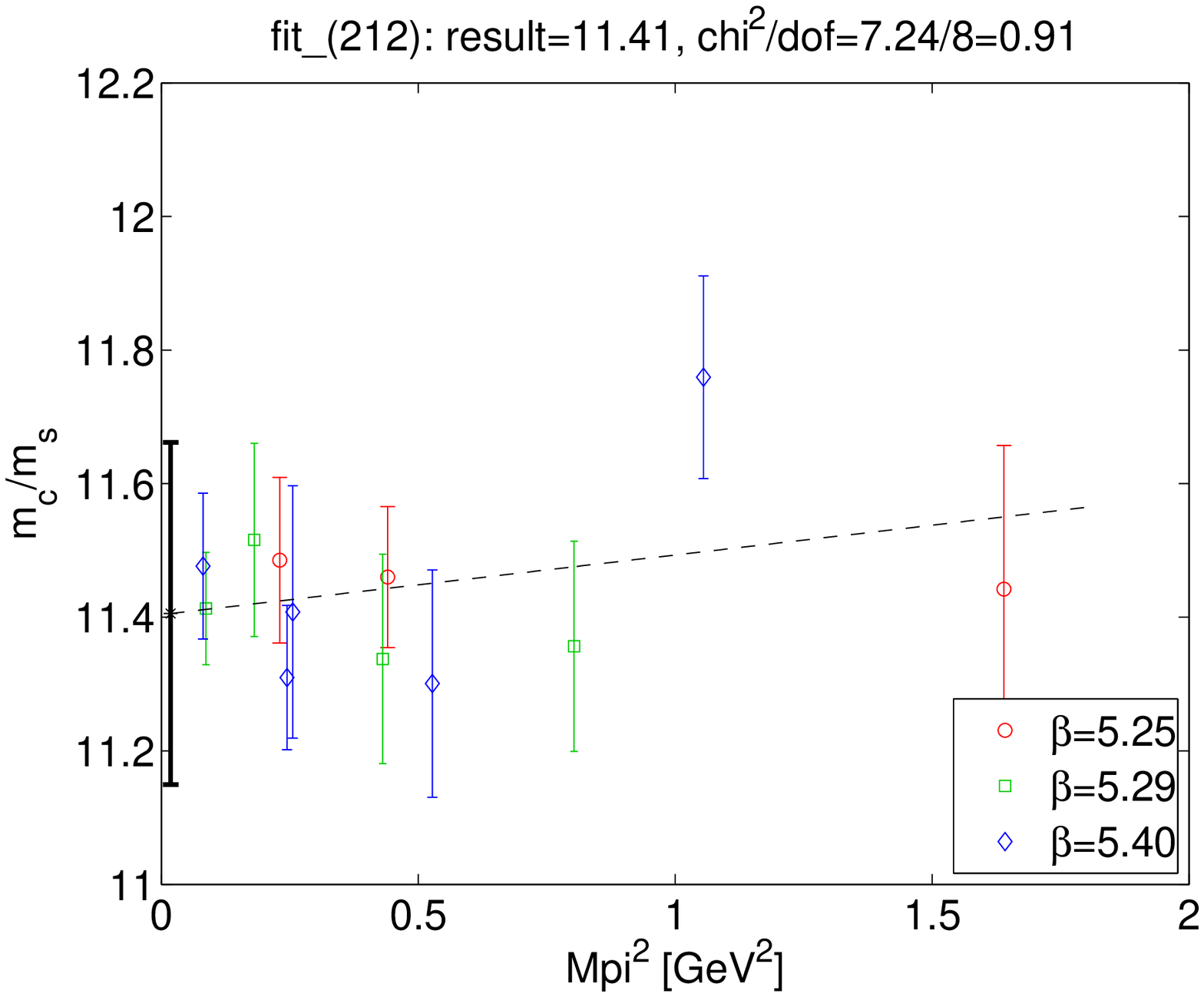}
\includegraphics[width=9cm]{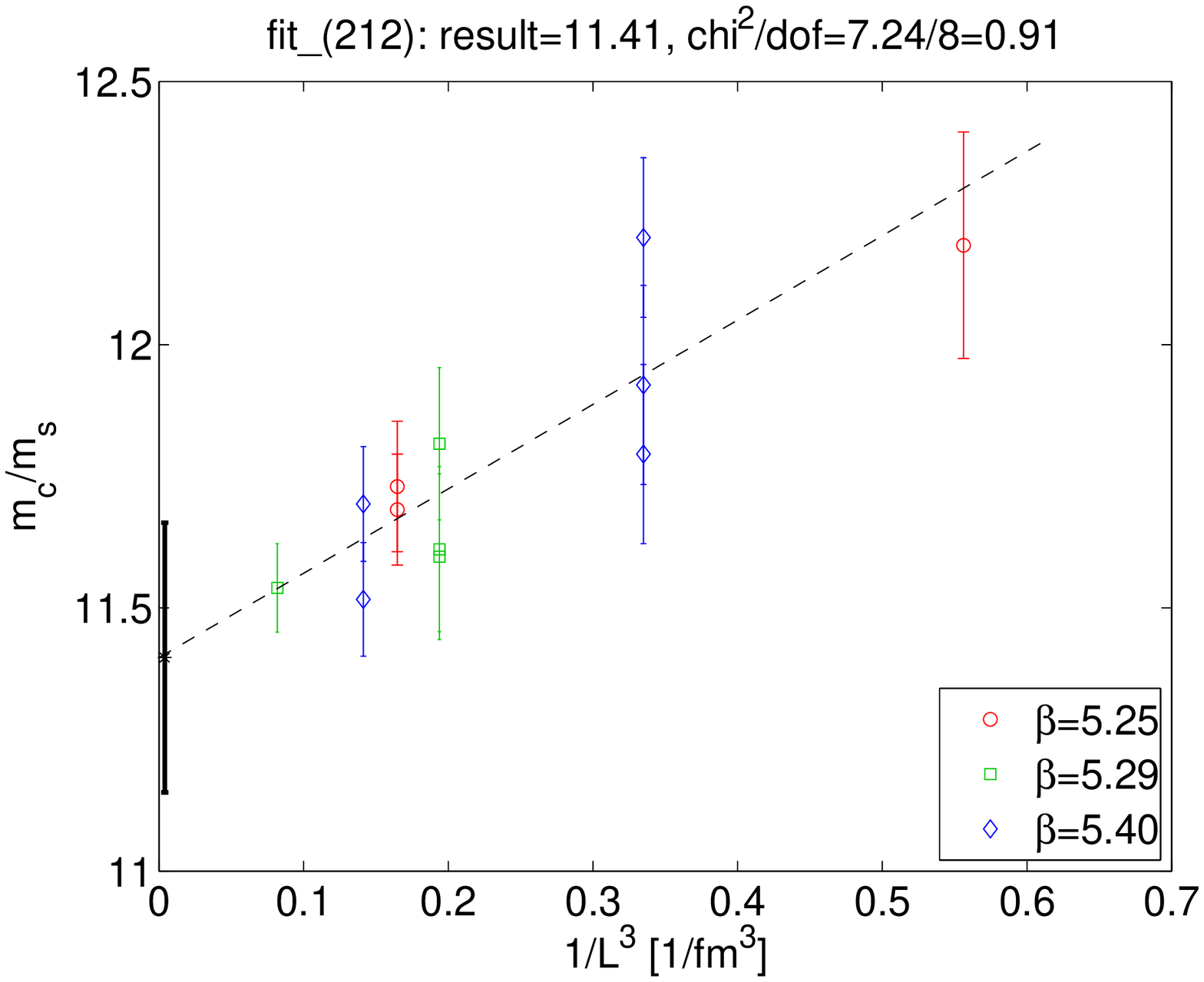}
%%%
%\includegraphics[width=9cm]{figures_v2/combiextrapolation_fit6_a.pdf}
%\includegraphics[width=9cm]{figures_v2/combiextrapolation_fit6_b.pdf}
%\includegraphics[width=9cm]{figures_v2/combiextrapolation_fit6_c.pdf}
%%%
\caption{\label{fig:extrapolation}\sl
One of the 8 global fits, namely $r^{(2,1,2)}$ with $O(a^2)$, $O(\Mpi^2)$,
$O(1/L^3)$ correction terms, for the joint extrapolation to zero lattice
spacing, physical pion mass, and infinite volume.}
\end{figure}

To illustrate the procedure we present one of the 8 global fits --~the
$(i,j,k)\!=\!(2,1,2)$ variety with $O(a^2)$, $O(\Mpi^2)$ and $O(1/L^3)$
terms~-- in Fig.\,\ref{fig:extrapolation}.
The data have been shifted by the effect of those terms which are not on
display.
For instance, in the continuum extrapolation panel
\bea
\mr{plotdata}(a)&=&
\mr{data}(a,\Mpi,L)-\mr{fit}(a,\Mpi,L)+\mr{fit}(a,\Mpi^\mr{phys},\infty)
\eea
is shown as a function of $a^2$, while in the pion mass extrapolation panel the
last term reads ``$\mr{fit}(0,\Mpi,\infty)$'', and in the infinite volume
extrapolation panel it is ``$\mr{fit}(0,\Mpi^\mr{phys},L)$''.
Note that this affects only the presentation, not the final result
(\ref{result}).

To test whether our assessment of systematic uncertainties is true and fair, we
apply the same analysis procedure to the observable
$O_4\!=\!M_\ph^2/(M_{D_s^*}^2\!-\!M_{D_s}^2)$.
This gives $1.79(08)(12)$, which agrees perfectly with the physical value
$1.7707$ \cite{Nakamura:2010zzi}.
This supports the view that our analysis procedure yields reliable estimates
of the uncertainties in (\ref{result}).

%%%%%%%%%%%%%%%%%%%%%%%%%%%%%%%%%%%%%%%%%%%%%%%%%%%%%%%%%%%%%%%%%%%%%%%%%%%%%%%

\section{From quark mass ratios to individual masses}

%%%%%%%%%%%%%%%%%%%%%%%%%%%%%%%%%%%%%%%%%%%%%%%%%%%%%%%%%%%%%%%%%%%%%%%%%%%%%%%

To give the reader an idea of what can be done with our result (\ref{result}),
we combine it with an aggregate value of $m_c$ to obtain an estimate of $m_s$.
For $m_c$ several precise results are available, which use either sum rule
techniques or perturbative estimates of moments of current correlators.
By contrast, computing $m_s$ directly on the lattice involves renormalization
factors like the factor $Z_A/Z_P$ in (\ref{awi_bhattacharya}) whose
non-perturbative determination is technically quite demanding.
Therefore, computing $m_s$ via (\ref{result}) from $m_c$ offers the
possibility to check the current best calculations of $m_s$ (see
\cite{Colangelo:2010et} for an overview) without recurrence to $Z$-factors
\cite{Davies:2009ih}.

We now collect the current best estimates of the charm mass, which have a
1-2\% error.
%%% $m_c(3\GeV)=1.008(26)\GeV$ from 1009.4325 by Bodenstein et al.\\
% $m_c(3\GeV)=0.987(09)\GeV$ from 1102.3835 by Bodenstein et al.\\
%%% Chetyrkin et al PRD 80, 074010 (2009) claims less than 1\% accuracy.\\
%%% A. Hoang and M. Jamin, Nucl. Phys. B 594, 127 (2004) claim what ?\\
% $m_c(3\GeV)=0.986(13)\GeV$ from 1010.6157 by Chetyrkin et al.
% $m_c(m_c)=1.277(26)\GeV$ from 1102.2264 by Dehnadi et al.\\
% $m_c(3\GeV)=0.986(6)\GeV$ from 1004.4285 by McNeile et al [HPQCD].
The first result $m_c(3\GeV)\!=\!0.986(6)\GeV$ \cite{McNeile:2010ji} is based
on the current correlator method on the lattice.
The remaining ones are based on sum rules and experimental electron-positron
annihilation cross section data, namely
$m_c(3\GeV)\!=\!0.986(13)\GeV$ \cite{Chetyrkin:2010ic},
$m_c(m_c)\!=\!1.277(26)\GeV$ \cite{Dehnadi:2011gc}, and
$m_c(3\GeV)\!=\!0.987(09)\GeV$ \cite{Bodenstein:2011ma}, respectively (for
an examination of the uncertainties involved see in
particular \cite{Dehnadi:2011gc}).
Through standard 4-loop $\MSbar$ running, these results can be evolved to
the common scale $\mu\!=\!2\GeV$, where they read
$m_c(2\GeV)\!=\!1.092(7),1.092(14),1.096(22),1.093(10)$ GeV, respectively.
A straight mean of the central values and of the systematic uncertainties
yields the conservative average $m_c(\MSbar,2\GeV)\!=\!1.093(13)\GeV$
\cite{McNeile:2010ji,Chetyrkin:2010ic,Dehnadi:2011gc,Bodenstein:2011ma}.

%%% mc_cen:=1093.: mc_sta:=0.000: mc_sys:=13.00: # v1
%%% cs_cen:=11.34: cs_sta:=00.40: cs_sys:=00.21: # v1
%%% sl_cen:=27.53: sl_sta:=00.20: sl_sys:=00.08: # v1
%%% la_cen:=0.381: la_sta:=0.005: la_sys:=0.027: # v1

%%% mc_cen:=1093.: mc_sta:=0.000: mc_sys:=13.00: # v2
%%% cs_cen:=11.27: cs_sta:=00.30: cs_sys:=00.26: # v2
%%% sl_cen:=27.53: sl_sta:=00.20: sl_sys:=00.08: # v2
%%% la_cen:=0.381: la_sta:=0.005: la_sys:=0.027: # v2

%%% ms_cen:=mc_cen/cs_cen;
%%% ms_sta:=sqrt((mc_sta/mc_cen)^2+(cs_sta/cs_cen)^2)*ms_cen;
%%% ms_sys:=sqrt((mc_sys/mc_cen)^2+(cs_sys/cs_cen)^2)*ms_cen;

%%% ml_cen:=ms_cen/sl_cen;
%%% ml_sta:=sqrt((ms_sta/ms_cen)^2+(sl_sta/sl_cen)^2)*ml_cen;
%%% ml_sys:=sqrt((ms_sys/ms_cen)^2+(sl_sys/sl_cen)^2)*ml_cen;

%%% mu_cen:=ml_cen*(1-la_cen);
%%% mu_sta:=sqrt((ml_sta/ml_cen)^2+(la_sta/(1-la_cen))^2)*mu_cen;
%%% mu_sys:=sqrt((ml_sys/ml_cen)^2+(la_sys/(1-la_cen))^2)*mu_cen;

%%% md_cen:=ml_cen*(1+la_cen);
%%% md_sta:=sqrt((ml_sta/ml_cen)^2+(la_sta/(1+la_cen))^2)*md_cen;
%%% md_sys:=sqrt((ml_sys/ml_cen)^2+(la_sys/(1+la_cen))^2)*md_cen;

Upon combining this input value with our result (\ref{result}) we arrive at the
estimate
\beq
m_s(\MSbar,2\GeV)=97.0(2.6)(2.5)\MeV
\label{indirect_s}
\eeq
which does not build on a renormalization factor.
At this point we may continue by using the ratios $m_s/m_{ud}=27.53(20)(08)$
and $(m_d\!-\!m_u)/(m_d\!+\!m_u)=0.381(05)(27)$ by the
Budapest-Marseille-Wuppertal collaboration \cite{Durr:2010vn,Durr:2010aw},
where $m_{ud}\!\equiv\!(m_u\!+\!m_d)/2$, to end up with
\beq
m_{ud}=3.52(10)(09)\MeV\;,\quad
m_u=2.18(06)(11)\MeV\;,\quad
m_d=4.87(14)(16)\MeV
\;.
\label{indirect_l}
\eeq
Still, the precision reached is competitive in view of the global averages
given in \cite{Colangelo:2010et}.

This concludes our illustration how the light quark masses can be obtained
without recurrence to renormalization factors, at the price of including
perturbative information.

%%%%%%%%%%%%%%%%%%%%%%%%%%%%%%%%%%%%%%%%%%%%%%%%%%%%%%%%%%%%%%%%%%%%%%%%%%%%%%%

\section{Summary}

%%%%%%%%%%%%%%%%%%%%%%%%%%%%%%%%%%%%%%%%%%%%%%%%%%%%%%%%%%%%%%%%%%%%%%%%%%%%%%%

The goal of this note has been to calculate the ratio $m_c/m_s$, using our
relativistic fermion action \cite{Durr:2010ch} in the valence sector, with
a controlled extrapolation to zero lattice spacing, to physical sea pion mass
and infinite box volume.
The only systematic effect which is not controlled is the quenching of the
strange and/or charm quark, but this is the case in other state-of-the-art
calculations \cite{Davies:2009ih,Blossier:2010cr}, too, and there are good
reasons to believe that the effect is negligible on the scale of the error
in (\ref{result}) (cf.\ the discussion in \cite{Colangelo:2010et}).

%%% 1.093/0.0955=11.44 from state-of-the-art calculations [mc_aggregate/ms_BMW]

%%% abs(11.27-11.85)/sqrt(0.30^2+0.26^2+0.16^2); ==> 1.36 sigma
%%% abs(11.27-12.00)/sqrt(0.30^2+0.26^2+0.30^2); ==> 1.47 sigma

%%% (11.85/0.16^2+12.0/0.30^2)/(1.0/0.16^2+1.0/0.30^2); --> 11.89 (aggreg. cen)
%%% sqrt(1.0/(1.0/0.16^2+1.0/0.30^2));                  -->  0.14 (aggreg. err)
%%% abs(11.27-11.89)/sqrt(0.30^2+0.26^2+0.14^2); ==> 1.47 sigma

Our result (\ref{result}) is consistent with the values
$m_c/m_s\!=\!11.85(16)$ by HPQCD \cite{Davies:2009ih} and $12.0(3)$ by ETM
\cite{Blossier:2010cr} (note that the spread among the entries in their
Tab.\,7 has not been propagated into their final error), with a slight
tension at the level of $1.36\sigma$ and $1.47\sigma$, respectively.
Though nominally less precise, our result serves as an important benchmark,
since our formulation bears the unique feature that it is free of any
lattice-induced isospin (or taste) breaking.
The relatively mild slope in $\al_\mr{s}a$ or $a^2$ as determined by our global
fits and the small overall spread among the entries in the $O_3\!=\!m_c/m_s$
column of Tab.\,\ref{tab:results} support the view that the formulation
\cite{Durr:2010ch} entails small cut-off effects up to the region of the
physical charm quark mass.

For illustration we combine our ratio (\ref{result}) with an average of $m_c$
from \cite{McNeile:2010ji,Chetyrkin:2010ic,Dehnadi:2011gc,Bodenstein:2011ma} to
obtain the value (\ref{indirect_s}) of $m_s$.
While there are results on $m_s$ with a higher claimed precision (see e.g.\
\cite{Colangelo:2010et} for a review), our computation is the only one which
avoids both $Z$-factors and unphysical isospin breaking effects, and this
renders the result particularly robust and reliable.

\bigskip\noindent{\bf Acknowledgments}: We thank the QCDSF collaboration for
allowing us to use their $\Nf\!=\!2$ configurations
\cite{Gockeler:2006jt,Gockeler:2006vi,Bietenholz:2010az,Collins:2011mk}
and the ILDG for making them available \cite{Yoshie:2008aw}.
We thank Thomas Lippert for support, and Zolt\'an Fodor and Stefan Sint for
discussion.
We acknowledge partial support in SFB/TR-55.
CPU resources on JUROPA were provided by Forschungszentrum J\"ulich GmbH.

%%%%%%%%%%%%%%%%%%%%%%%%%%%%%%%%%%%%%%%%%%%%%%%%%%%%%%%%%%%%%%%%%%%%%%%%%%%%%%%

% NOTE: \cite{Gockeler:2006vi} not needed, once extra columns in Tab.1 dropped.

\clearpage


\begin{thebibliography}{99}

%%%%%%%%%%%%%%%%%%%%%%%%%%%%%%%%%%%%%%%%%%%%%%%%%%%%%%%%%%%%%%%%%%%%%%%%%%%%%%%

\itemsep-2pt

\bibitem{Nakamura:2010zzi}
  K.~Nakamura {\it et al.} [Particle Data Group],
  %``Review of particle physics,''
  J.\ Phys.\ G {\bf 37}, 075021 (2010).
  %%CITATION = JPHGB,G37,075021;%%
\bibitem{Colangelo:2010et}
  G.~Colangelo {\it et al.} [FLAG],
  %``Review of lattice results concerning low energy particle physics,''
  Eur.\ Phys.\ J.\ C {\bf 71}, 1695 (2011) [arXiv:1011.4408].
  %%CITATION = EPHJA,C71,1695;%%
\bibitem{Rosner:2010ak}
  J.~L.~Rosner and S.~Stone,
  %``Leptonic Decays of Charged Pseudoscalar Mesons,''
  arXiv:1002.1655 [hep-ex].
  %%CITATION = ARXIV:1002.1655;%%

\bibitem{Davies:2009ih}
  C.~T.~H.~Davies {\it et al.} [HPQCD],
  %``Precise Charm to Strange Mass Ratio and Light Quark Masses from Full
  %Lattice QCD,''
  Phys.\ Rev.\ Lett.\ {\bf 104}, 132003 (2010) [arXiv:0910.3102].
  %%CITATION = PRLTA,104,132003;%%
\bibitem{Blossier:2010cr}
  B.~Blossier {\it et al.}  [ETM Collab.],
  %``Average up/down, strange and charm quark masses with Nf=2 twisted mass
  %lattice QCD,''
  Phys.\ Rev.\ D {\bf 82}, 114513 (2010), [arXiv:1010.3659].
  %%CITATION = PHRVA,D82,114513;%%

\bibitem{Durr:2010ch}
  S.~Durr and G.~Koutsou,
  %``Brillouin improvement for Wilson fermions,''
  Phys.\ Rev.\ D {\bf 83}, 114512 (2011) [arXiv:1012.3615].
  %%CITATION = PHRVA,D83,114512;%%

\bibitem{Bhattacharya:2005rb}
  T.~Bhattacharya {\it et al.},
  %``Improved bilinears in lattice QCD with non-degenerate quarks,''
  Phys.\ Rev.\ D {\bf 73}, 034504 (2006) [hep-lat/0511014].
  %%CITATION = PHRVA,D73,034504;%%

\bibitem{Gockeler:2006jt}
  M.~Gockeler {\it et al.} [QCDSF Collab.],
  %``Estimating the unquenched strange quark mass from the lattice axial Ward
  %identity,''
  Phys.\ Rev.\ D {\bf 73}, 054508 (2006) [hep-lat/0601004].
  %%CITATION = PHRVA,D73,054508;%%
\bibitem{Gockeler:2006vi}
  M.~Gockeler {\it et al.} [QCDSF Collab.],
  %``Simulating at realistic quark masses: Light quark masses,''
  PoS (LAT2006) 160 (2006) [hep-lat/0610071].
  %%CITATION = POSCI,LAT2006,160;%%
\bibitem{Bietenholz:2010az}
  W.~Bietenholz {\it et al.} [QCDSF Collab.],
  %``Pion in a Box,''
  Phys.\ Lett.\ B {\bf 687}, 410 (2010) [arXiv:1002.1696].
  %%CITATION = PHLTA,B687,410;%%
\bibitem{Collins:2011mk}
  S.~Collins {\it et al.} [QCDSF Collab.],
  %``Dirac and Pauli form factors from lattice QCD,''
  Phys.\ Rev.\ D\ {\bf 84}, 074507  (2011) [arXiv:1106.3580].
  %%CITATION = PHRVA,D84,074507;%%

\bibitem{Yoshie:2008aw}
  T.~Yoshi\'e,
  %``Making use of the International Lattice Data Grid,''
  PoS (LAT2008) 019 (2008) [arXiv:0812.0849].
  %%CITATION = POSCI,LATTICE2008,019;%%

\bibitem{arXiv:1111.2577}
  S.~Durr and G.~Koutsou,
  %``$m_c/m_s$ with Brillouin fermions,''
  PoS (LAT2011) 230 (2011) [arXiv:1111.2577].
  %%CITATION = ARXIV:1111.2577;%%

\bibitem{McNeile:2010ji}
  C.~McNeile {\it et al.} [HPQCD Collab.],
  %C.~T.~H.~Davies, E.~Follana, K.~Hornbostel and G.~P.~Lepage,
  %``High-Precision c and b Masses, and QCD Coupling from Current-Current
  %Correlators in Lattice and Continuum QCD,''
  Phys.\ Rev.\ D {\bf 82}, 034512 (2010) [arXiv:1004.4285].
  %%CITATION = PHRVA,D82,034512;%%
\bibitem{Chetyrkin:2010ic}
  K.~Chetyrkin {\it et al.},
  %J.~H.~Kuhn, A.~Maier, P.~Maierhofer, P.~Marquard, M.~Steinhauser and
  %C.~Sturm, ``Precise Charm- and Bottom-Quark Masses: Theoretical and
  %Experimental Uncertainties,''
  arXiv:1010.6157 [hep-ph].
  %%CITATION = ARXIV:1010.6157;%%
\bibitem{Dehnadi:2011gc}
  B.~Dehnadi, A.~H.~Hoang, V.~Mateu and S.~M.~Zebarjad,
  %``Charm Mass Determination from QCD Charmonium Sum Rules at Order
  %alpha_s^3,''
  arXiv:1102.2264 [hep-ph].
  %%CITATION = ARXIV:1102.2264;%%
\bibitem{Bodenstein:2011ma}
  S.~Bodenstein {\it et al.},
  %J.~Bordes, C.~A.~Dominguez, J.~Penarrocha and K.~Schilcher,
  %``QCD sum rule determination of the charm-quark mass,''
  Phys.\ Rev.\  D {\bf 83}, 074014 (2011) [arXiv:1102.3835].
  %%CITATION = PHRVA,D83,074014;%%

\bibitem{Durr:2010vn}
  S.~Durr {\it et al.} [BMW Collab.],
  %``Lattice QCD at the physical point: light quark masses,''
  Phys.\ Lett.\ B {\bf 701}, 265 (2011), [arXiv:1011.2403].
  %%CITATION = PHLTA,B701,265;%%
\bibitem{Durr:2010aw}
  S.~Durr {\it et al.} [BMW Collab.],
  %``Lattice QCD at the physical point: Simulation and analysis details,''
  JHEP\ {\bf 1108}, 148 (2011) [arXiv:1011.2711].
  %%CITATION = JHEPA,1108,148;%%

\end{thebibliography}
\end{document}